\documentclass[conference]{IEEEtran}
\input{psfig.sty}
\usepackage{subfigure}
\usepackage{multirow}
\usepackage{psfig}
\usepackage{graphicx}
\usepackage{amsmath}
\usepackage{epsfig}
\usepackage{amsfonts}
\usepackage{amssymb}
\usepackage{amsthm}
\usepackage{algorithm}
\usepackage{algorithmic}
\usepackage[usenames,dvipsnames]{pstricks}
\usepackage{arydshln}

\newcommand{\argmin}{\mathop{\text{argmin}}}

\newcommand{\vs}{{\bf s}}
\newcommand{\vx}{{\bf x}}
\newcommand{\vy}{{\bf y}}

\newcommand{\vn}{{\bf n}}

\newcommand{\pp}{{\bf P}}
\newcommand{\pq}{{\bf Q}}
\newcommand{\mh}{{\bf H}}

\newcommand{\mg}{{\bf G}}

\newcommand{\ma}{{\bf A}}
\newcommand{\mb}{{\bf B}}
\newcommand{\md}{{\bf D}}

\newcommand{\sm}{{\mathbb S}_{n_t,{\mathbb A}}}
\newcommand{\sv}{{\mathbb S}_{n_p,{\mathbb A}}}
\newcommand{\sa}{{\mathbb A}}

\newcommand{\Define}{\triangleq}

\begin{document}
\twocolumn

\title{{\huge 
Precoder Index Modulation
}}

\author{T. Lakshmi Narasimhan and A. Chockalingam \\
Department of ECE, Indian Institute of Science, Bangalore 560012, India
}

\IEEEaftertitletext{\vspace{-0.6\baselineskip}}
\maketitle
\begin{abstract}
Index modulation, where information bits are conveyed through antenna 
indices (spatial modulation) and subcarrier indices (subcarrier index
modulation) in addition to information bits conveyed through conventional 
modulation symbols, is getting increased research attention. In this paper, 
we introduce {\em precoder index modulation}, where information bits are 
conveyed through the choice of a precoder matrix at the transmitter from 
a set of pre-determined pseudo-random phase precoder (PRPP) matrices. 
Combining precoder index modulation (PIM) and spatial modulation (SM), 
we introduce a PIM-SM scheme which conveys information bits through both 
antenna index as well as precoder index. Spectral efficiency (in bits per 
channel use) and bit error performance of these index modulation schemes 
are presented.
\end{abstract}

\section{Introduction}
\label{sec1}
In conventional modulation schemes, information bits are conveyed in 
a given channel use through the transmission of a modulation symbol 
chosen from a pre-determined modulation alphabet (e.g., QAM, PSK) 
\cite{proakis}. Recently, modulation techniques that convey additional 
information bits through antenna indices and subcarrier indices have 
generated increased research interest. One such index modulation scheme 
that is popularly researched in recent times is {\em spatial modulation} 
(see \cite{sm},\cite{book} and the references therein), which is a 
multi-antenna modulation scheme that uses antenna indexing. Another 
such scheme is {\em subcarrier index modulation}, which is a 
multi-carrier modulation scheme that uses subcarrier indexing 
\cite{sub_index1},\cite{sub_index2}.

Spatial modulation (SM) uses multiple transmit antennas but only one 
transmit radio frequency (RF) chain. In a given channel use, only 
one among all the transmit antennas will be activated and the others 
will remain silent. The activated antenna will carry a conventional 
modulation symbol. The antenna to be activated is chosen based on 
information bits. Therefore, the index of the activated antenna 
conveys additional information bits. If $n_t$ is the number of 
transmit antennas and $\sa$ is the modulation alphabet, the 
number of bits conveyed per channel use in SM is given by
$\lfloor \log_2|{\mathbb A}|\rfloor + \lfloor \log_2n_t\rfloor$.
Similarly, subcarrier index modulation refers to a multi-carrier 
modulation scheme, where information bits are conveyed through the 
indices of active subcarriers \cite{sub_index1},\cite{sub_index2}. 
An advantage of index modulation over conventional modulation is 
that, to achieve a certain spectral efficiency, index modulation can 
use a smaller-sized QAM/PSK alphabet compared to that needed in 
conventional modulation. This, in turn, can lead to SNR gains (for a 
given probability of error performance) in favor of index modulation 
\cite{marco},\cite{tln}. 

{\em Pseudo-random phase precoding (PRPP):} 
The link reliability in single-input-single-output (SISO) fading channels 
is poor due to lack of diversity. One way to improve the link reliability 
is to get diversity gains through the use of multiple antennas. Diversity
gains can be achieved even in single-antenna systems using rotation coding 
\cite{tse} or transmit power control \cite{vinod}. Transmit power control 
requires channel state information at the transmitter (CSIT). Whereas, 
rotation coding does not require CSIT. The idea in rotation coding is to 
use multiple channel uses and precode the transmit symbol vector using a 
phase precoder matrix without requiring more slots than the number of 
symbols precoded. A $2\times 2$ phase precoder matrix with optimized 
phases is shown to achieve a diversity gain of two in SISO fading channels 
\cite{tse}. In \cite{ramesh}, the rotation coding idea has been exploited
for large precoder sizes. Instead of using optimized phases in the precoder
matrix (solving for optimum phases for large precoder sizes is difficult), 
pseudo-random phases are used. Also, the issue of detection complexity at 
the receiver for large precoder sizes has been addressed by using the 
low complexity likelihood ascent search (LAS) algorithm in \cite{lmimo1}. 
It has been shown that with pseudo-random phase precoding (PRPP) and LAS
detection, near-exponential diversity is achieved in a SISO fading channel
for large precoder sizes (e.g., $300\times 300$ precoder matrix).

{\em PRPP-SM:}
In \cite{prpp_sm1},\cite{prpp_sm2}, a PRPP-SM scheme which simultaneously
exploits the diversity advantage of PRPP and the SNR advantage of SM is 
proposed. The PRPP-SM scheme precodes both the modulation bits and the 
antenna index bits using pseudo-random phases. The PRPP-SM scheme has
been shown to outperforms both the PRPP scheme without SM and the SM 
scheme without PRPP.

In this paper, we introduce the idea of index modulation applied to the 
choice of precoder matrix at the transmitter, which we refer to as 
{\em precoder index modulation}. In the precoder index modulation (PIM) 
scheme, additional information bits are conveyed through the choice of 
a precoding matrix from a set of pre-determined PRPP matrices. We
also combine the PIM and PRPP-SM schemes, and propose a PIM-SM scheme 
which conveys bits through both antenna index as well as precoder index.
The rest of the paper is organized as follows. The PRPP, SM, and PRPP-SM
schemes are introduced in Section \ref{sec2}. The PIM and PIM-SM schemes 
are presented in Section \ref{sec3}. Conclusions are presented in Section 
\ref{sec4}.

\begin{figure}[t]
\centering
\includegraphics[height=0.75in,width=3.4in]{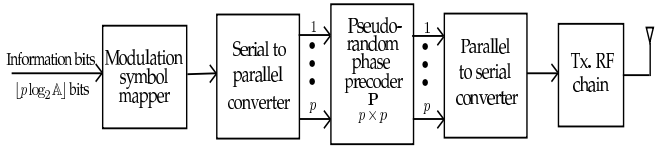}
\caption{PRPP transmitter.}
\vspace{-4mm}
\label{fig_prpp}
\end{figure}

\section{PRPP, SM, and PRPP-SM schemes}
\label{sec2}
In this section, we briefly introduce PRPP, SM, and PRPP-SM schemes.

\subsection{PRPP scheme}
\label{sec2a}
\begin{figure}[h]
\centering
\vspace{-2mm}
\includegraphics[height=2.5in,width=3.5in]{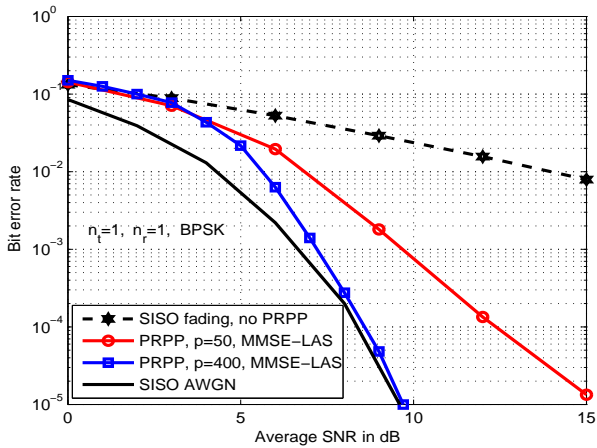}
\caption{Performance of PRPP in SISO fading with $p=50,400$, BPSK. 
and LAS detection.}
\vspace{-4mm}
\label{fig_prpp_perf}
\end{figure}

Figure \ref{fig_prpp} shows the PRPP transmitter. The PRPP transmitter
takes $p$ modulated 
symbols and forms the symbol vector $\vs \in \sa^p$, where $\sa$ is the 
modulation alphabet. The symbol vector $\vs$ is then precoded using a 
$p\times p$ precoding matrix $\pp$ to get the transmit vector 
$\pp\mathbf s$. The $(r,c)$th entry of the precoder matrix $\pp$ 
is $\frac{1}{\sqrt{p}}e^{j\theta_{r,c}}$, where the phases 
$\{\theta_{r,c}\}$ are generated using a pseudo-random sequence 
generator. The seed of this random number generator is pre-shared 
among the transmitter and receiver. The precoded sequence $\pp\vs$ is 
transmitted through the channel, which is assumed to be frequency-flat 
fading. The channel fade coefficients are assumed to be i.i.d from one 
channel use to the other. At the receiver, after $p$ channel uses, the 
received symbols are accumulated to form the $p\times 1$ received
vector $\vy$, given by
\begin{eqnarray}
\vy & = &\md\pp\mathbf s + \vn \nonumber \\
& = & \mg\mathbf s + \vn,
\end{eqnarray}
where $\md=\text{diag}\{h_{(1)}\, h_{(2)}\, \cdots\, h_{(p)}\}$, 
$\mg=\md\pp$, $h_{(i)}$s are i.i.d. complex Gaussian fade coefficients 
with zero mean and unit variance, and $\vn$ is the $p\times 1$ noise 
vector $[n_{(1)}\, n_{(2)}\, \cdots\, n_{(p)}]^T$
whose entries are distributed as ${\mathbb C}{\mathcal N}(0,\sigma^2)$. 
The entries of the matrix $\mg$ are uncorrelated and 
$\lVert \md\rVert_F=\lVert \mg \rVert_F$. This creates a $p\times p$ 
virtual MIMO system. It has been shown in \cite{ramesh} that as the 
precoder size becomes large (e.g., $p=300$) the performance of PRPP 
in SISO fading, using the likelihood ascent search (LAS) detection 
algorithm in \cite{lmimo1} with MMSE initial solution, approaches 
exponential diversity performance (i.e., close to SISO AWGN performance). 
This point is illustrated in Fig.  \ref{fig_prpp_perf} which shows the 
performance of PRPP with BPSK modulation for $p=50$ and 400 in SISO
fading channels.

\subsection{SM scheme}
\label{sec2b}
\begin{figure}[h]
\centering
\includegraphics[height=1.1in,width=1.5in]{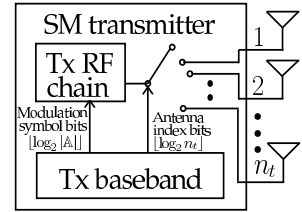}
\caption{SM transmitter with $n_t$ antennas and one transmit RF chain.}
\vspace{-2mm}
\label{sm_fig}
\end{figure}

The SM scheme uses $n_t$ transmit antennas but only one transmit RF chain
as shown in Fig. \ref{sm_fig}. The number of transmit RF chains, $n_{rf}=1$. 
In a given channel use, the transmitter selects one of its $n_t$ transmit 
antennas, and transmits a modulation symbol from the alphabet $\sa$ 
on the selected antenna. The number of bits transmitted per channel use 
through the modulation symbol is $\lfloor \log_2|{\mathbb A}| \rfloor$, 
and the number of bits conveyed per channel use through the index of the 
transmitting antenna is $\lfloor \log_2n_t \rfloor$. Therefore, a total 
of $\lfloor \log_2|{\mathbb A}|\rfloor + \lfloor \log_2n_t\rfloor$ bits 
per channel use (bpcu) is conveyed. For e.g., in an SM system with $n_t=2$ 
and 8-QAM, the spectral efficiency is 4 bpcu. 

\begin{figure*}[t]
\centering
\includegraphics[height=2in,width=7in]{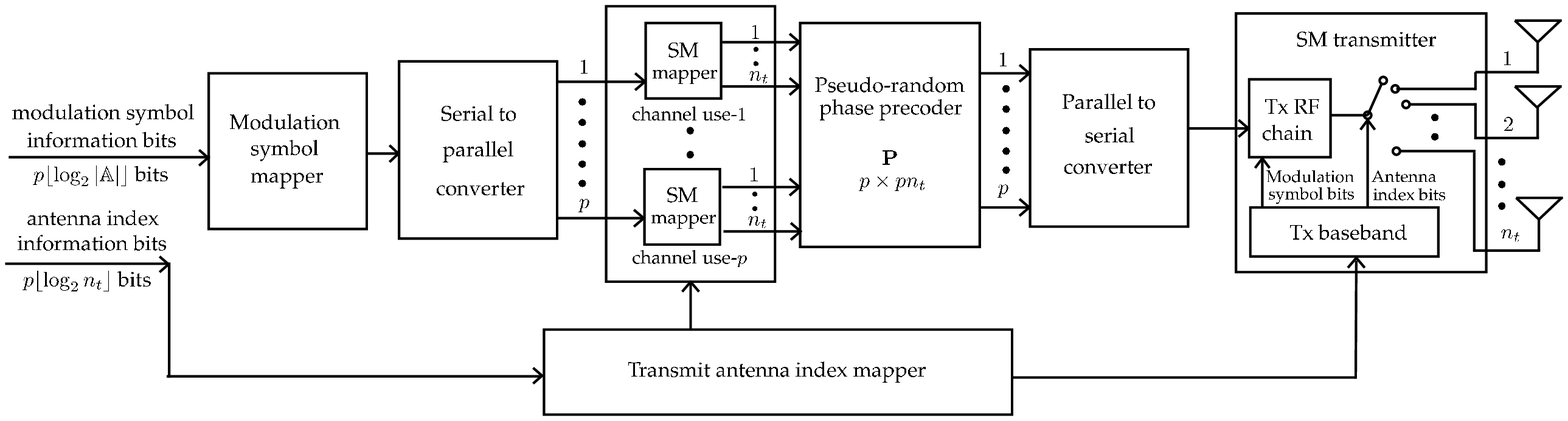}
\caption{PRPP-SM scheme.}
\vspace{0mm}
\label{sys}
\end{figure*}

The SM alphabet set for a fixed $n_t$ and $\sa$ is given by 
\begin{eqnarray}
\sm = 
\big \{ {\bf x}_{j,l}:j=1,\cdots,n_t, \ \ l=1,\cdots,|{\mathbb A}| \big \}, \nonumber \\ 
\mbox{s.t.} \ \ {\bf x}_{j,l} = 
[0,\cdots,0,\hspace{-4mm}\underbrace{s_{l}}_{{\scriptsize{\mbox{$j$th coordinate}}}}\hspace{-3.5mm},0,\cdots,0]^T, \ \ x_l \in \mathbb{A}.  
\label{eqy1}
\end{eqnarray}
For e.g., for $n_t=2$ and 4-QAM, ${\mathbb S}_{n_t,{\mathbb A}}$ is 
given by  
{\small
\begin{eqnarray}
\hspace{-4mm}
{\mathbb S}_{2,\mbox{{\tiny 4-QAM}}}  
\hspace{-3mm}&=&\hspace{-3mm}\Bigg\{ 
\begin{bmatrix} +1+j \\ 0 \end{bmatrix}, 
\begin{bmatrix} +1-j \\ 0 \end{bmatrix}, 
\begin{bmatrix} -1+j \\ 0 \end{bmatrix}, 
\begin{bmatrix} -1-j \\ 0 \end{bmatrix}, \nonumber \\ 
& & 
\begin{bmatrix} 0 \\ +1+j \end{bmatrix}, 
\begin{bmatrix} 0 \\ +1-j \end{bmatrix}, 
\begin{bmatrix} 0 \\ -1+j \end{bmatrix}, 
\begin{bmatrix} 0 \\ -1-j \end{bmatrix} 
\Bigg \}. 
\end{eqnarray}
}

\vspace{-2mm}
Let $\vx \in {\mathbb S}_{n_t,{\mathbb A}}$ denote the transmit vector.
Let $\mh \in \mathbb{C}^{n_r\times n_t}$ denote the channel gain matrix,
where $H_{i,j}$ denotes the channel gain from the $j$th transmit antenna 
to the $i$th receive antenna, assumed to be i.i.d complex Gaussian with
zero mean and unit variance. The received signal at the $i$th receive 
antenna is 
\begin{equation}
y_i = H_{i,j}x_l + n_i, \quad \quad i=1,\cdots,n_r,
\end{equation}
where $n_r$ is the number of receive antennas, $x_l$ is the $l$th symbol 
in ${\mathbb A}$ transmitted by the $j$th antenna, and $n_i$ is the noise
component. The signals received at all the receive antennas can be written 
in vector form as
\begin{eqnarray}
\vy& = & \mh\vx+\vn,
\label{sysmodel}
\end{eqnarray}
For this system model, the maximum likelihood (ML) detection rule is 
given by
\begin{equation} 
\label{ml1} 
\hat{\vx}=\argmin_{\vx\in \sm} \ \|\vy-\mh\vx\|^2.  
\end{equation}

\subsection{PRPP-SM scheme}
\label{sec2c}
The PRPP-SM scheme is an SM scheme whose modulation bits as well as
antenna index bits are precoded by a PRPP matrix 
\cite{prpp_sm1},\cite{prpp_sm2}. The PRPP-SM transmitter consists of 
$n_t$ transmit antennas and $n_{rf}=1$ transmit RF chains as shown in 
Fig. \ref{sys}. It takes $p$ modulated symbols and forms the symbol 
vector $\vx_s \in \sa^p$, where $\sa$ is the modulation alphabet.
A matrix $\ma$ of size $pn_t\times p$ denotes the transmit antenna 
activation pattern, such that $\ma\vx_s \in \sm^p$. The matrix $\ma$ 
consists of $p$ submatrices $\ma_{(i)}$, $i=1,\cdots,p$, each of size 
$n_t\times p$, such that 
$\ma=[\ma_{(1)}^T \, \ma_{(2)}^T \, \cdots \, \ma_{(p)}^T]^T$, and 
$\ma_{(i)}\vx_s \in \sm$. The submatrix $\ma_{(i)}$ is constructed as 
\begin{equation}
\ma_{(i)}=[{\bf 0}_{(1)} \, \cdots \, {\bf 0}_{(i-1)} \, {\bf a}_{(i)} \, {\bf 0}_{(i+1)} \, \cdots \, {\bf 0}_{(p)}],
\end{equation} 
where ${\bf 0}_{(k)}$ is a $n_t\times 1$ vector of zeroes, and 
${\bf a}_{(i)}$ is a $n_t\times 1$ vector constructed as
\begin{equation}
{\bf a}_{(i)}=[0 \, \cdots \, 0 \, \hspace{-2mm}\underbrace{1}_{\mbox{\tiny{$j_i$th coordinate}}} \hspace{-2mm} 0 \, \cdots \, 0]^T,
\end{equation}
where $j_i$ is the index of the active antenna during the $i$th channel 
use. For e.g., in a system with $n_t=2$ and $p=3$, to activate antennas
1, 2 and 1 in three consecutive channel uses, respectively, the matrix 
$\ma$ is given by
\begin{equation}
\ma \ = \ 
\begin{bmatrix}
\ma_{(1)}\\
\ma_{(2)}\\
\ma_{(3)}
\end{bmatrix}=
\begin{bmatrix}
1 & 0 & 0\\
0 & 0 & 0\\ \hdashline
0 & 0 & 0\\
0 & 1 & 0\\ \hdashline
0 & 0 & 1\\
0 & 0 & 0
\end{bmatrix}.
\label{eqx1}
\end{equation}
Note that the indices of the non-zero rows in matrix $\ma$ gives the 
support of the spatially modulated vector $\ma\vx_s\in\sm^p$. For e.g.,
in (\ref{eqx1}), the support given by $\ma$ is $\{1,4,5\}$.

The $\ma\vx_s$ vector is then precoded as $\pp\ma\vx_s$, using a rectangular 
precoder matrix $\pp$ of size $p\times pn_t$. The $(r,c)$th entry of the
$\pp$ matrix is $\frac{1}{\sqrt{p}}e^{j\theta_{r,c}}$, where the phases 
$\{\theta_{r,c}\}$ are generated using a pseudo-random sequence generator, 
whose seed is pre-shared among the transmitter and receiver. The output of 
the precoder is transmitted on the selected antenna in each channel use.

Let $n_r$ denote the number of receive antennas. The $pn_r\times 1$ 
received signal vector at the receiver is given by
\begin{eqnarray}
\vy &=&\md\ma\pp\ma\vx_s + \vn,
\end{eqnarray}
where $\md=\mbox{diag}\{\mh_{(1)}\, \mh_{(2)}\, \cdots\, \mh_{(p)}\}$, 
$\mh_{(i)}$ is the $n_r\times n_t$ channel matrix of the $i$th channel
use, the elements of $\mh_{(i)}$ are i.i.d. complex Gaussian with zero 
mean and unit variance, $\vn$ is the $pn_r\times 1$ noise vector, whose 
entries are distributed as $\mathbb{C}{\cal N}(0,\sigma^2)$. The ML 
detection rule for the PRPP-SM system is then given by
\begin{equation} 
\label{ml3} 
\{\hat{\vx}_s,\hat{\ma}\}=\argmin_{\vx_s\in \sa^p, \forall \ma} \ \|\vy-\md\ma\pp\ma\vx_s\|^2.  
\end{equation}
The indices of the non-zero rows in $\hat{\mathbf A}$ and the entries of 
$\hat{\mathbf x}_s$ are demapped to obtain the information bits.

\begin{figure}[t]
\includegraphics[height=2.75in,width=3.75in]{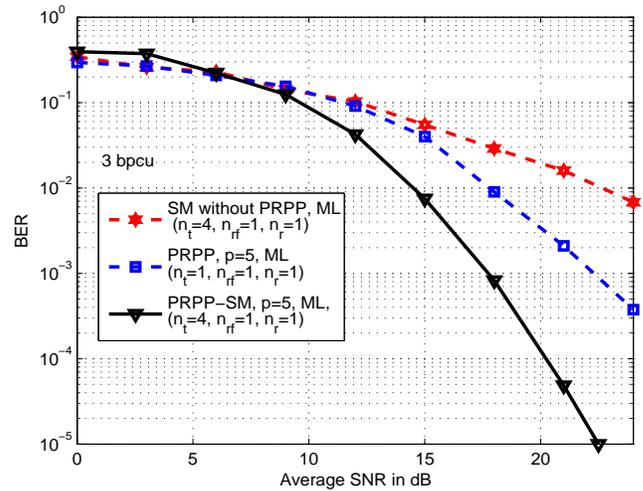}
\vspace{-6mm}
\caption{ML detection performance comparison of PRPP-SM scheme 
($n_t=4, n_{rf}=1, n_r=1, p=5$, BPSK) with SM scheme without PRPP 
($n_t=4, n_{rf}=1, n_r=1$, BPSK) and PRPP scheme without SM
($n_t=1, n_{rf}=1, n_r=1, p=5$, 8-QAM). Spectral efficiency is 3 bpcu.}
\label{prppsm}
\vspace{-5mm}
\end{figure}

For the same spectral efficiency, PRPP-SM scheme performs significantly 
better than SM scheme without PRPP and PRPP scheme without SM. This 
can be seen from Fig. \ref{prppsm}, which shows the bit error performance 
of PRPP-SM, PRPP, and SM schemes at the same spectral efficiency of 3 bpcu. 
For e.g., at $10^{-2}$ BER, the PRPP-SM scheme outperforms SM scheme without 
PRPP and PRPP scheme without SM by about 7 dB and 3 dB, respectively. This 
performance advantage in favor of the PRPP-SM scheme is due to the diversity 
gain offered by the pseudo-random phase precoding. 

\begin{figure*}
\hspace{-2mm}
\subfigure[PIM scheme]{
\includegraphics[height=1.5in,width=4.0in]{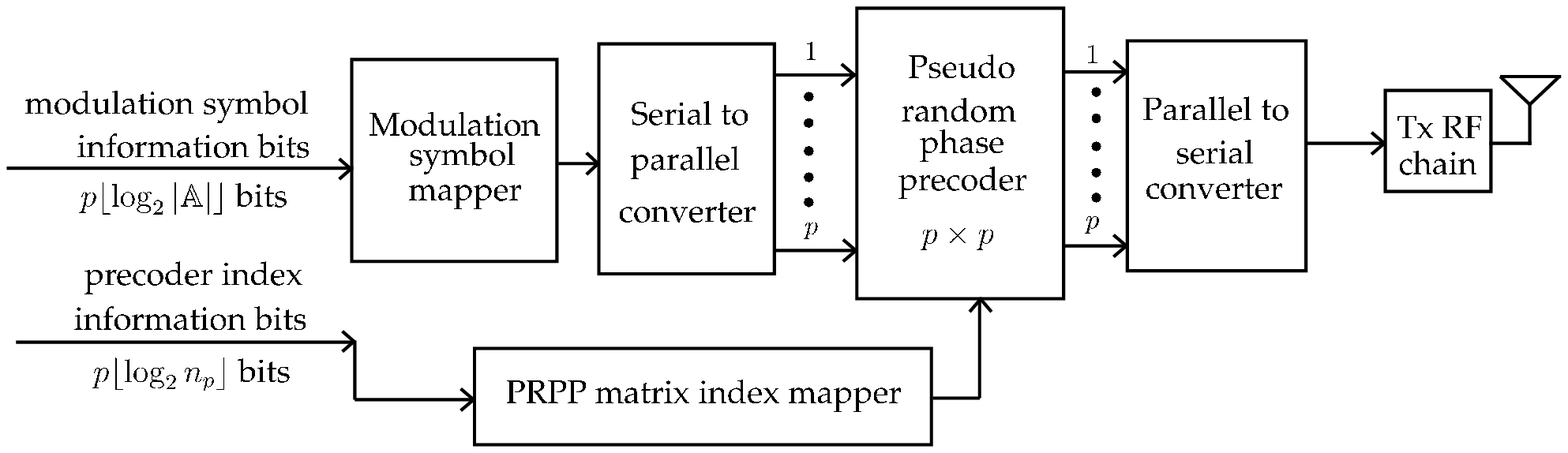}
\label{pimsys}}
\subfigure[PIM-SM scheme]{
\includegraphics[height=1.5in,width=3in]{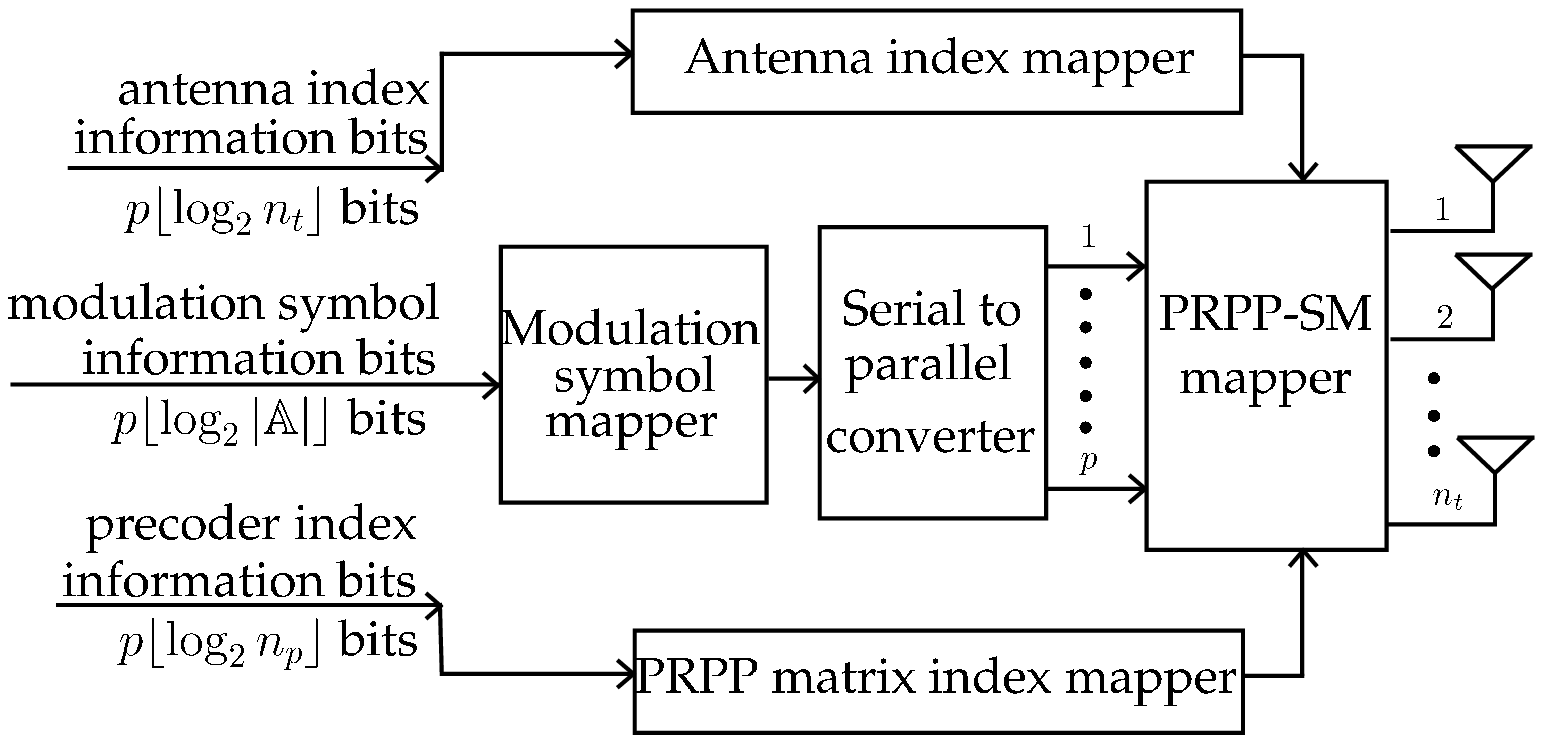}
\label{pimsmsys}}
\caption{Proposed PIM and PIM-SM schemes.}
\vspace{-5mm}
\end{figure*}

\section{Precoder index modulation}
\label{sec3}
In this section, we introduce precoder index modulation (PIM) which 
conveys information bits through the choice of a precoding matrix 
among a set of pre-determined precoder matrices.

\subsection{Precoder index modulation (PIM)}
\label{sec3a}
In the proposed PIM scheme, the transmitter has equal number of transmit 
antennas and RF chains. Consider the case of $n_t=n_{rf}=1$. The idea here 
is to have a collection of precoder matrices each of size $p\times p$ and 
choose one among these matrices to precode $p$ modulation symbols from
an alphabet $\sa$ in $p$ channel uses. Call this collection of matrices 
as `precoder set,' denoted by $\mathbb P$. Therefore, the number of bits 
conveyed per channel use through precoder indexing is 
$\frac{1}{p}\lfloor\log_2|{\mathbb P}|\rfloor$. The total number of bits 
per channel use (including precoder index bits and modulation symbol bits) 
is then given by
\[
\frac{1}{p}\Big(\lfloor\log_2|{\mathbb P}|\rfloor + p\lfloor\log_2|\sa|\rfloor\Big).
\]
{\em Construction of the precoder set:}
The precoder set $\mathbb P = \{\pp_1, \pp_2, \cdots, \pp_{|{\mathbb P}|}\}$ 
is constructed as follows.
Let $n$ denote the number of precoder index bits per channel use,
i.e., $n\Define \frac{1}{p}\lfloor\log_2|{\mathbb P}|\rfloor$.
Generate a PRPP matrix $\pq$ of size $p\times pn_p$, where $n_p=2^n$.
Note that the precoder set size $|{\mathbb P}|=(n_p)^p$. 
The matrix $\pq$ can be written as
\[
\pq = [\pq_1 \ \pq_2 \ \cdots \ \pq_p],
\]
where $\pq_i$'s are sub-matrices each of size $p\times n_p$. Now, the 
$p$ columns of a precoder matrix $\pp_j$ are obtained by drawing 
one column from each $\pq_i$, $i=1,2,\cdots,p$. Each of such draws 
form one precoder matrix. Since there are $n_p$ columns in each $\pq_i$, 
the number of such possible draws is $(n_p)^p$, which gives us all the 
matrices in the precoder set. For e.g., consider $p=2$ and $n_p=2$, and
\[
\pq=\left[ \begin{array}{cc|cc} q_{11} & q_{12} & q_{13} & q_{14}\\ 
q_{21} & q_{22} & q_{23} & q_{24} \end{array} \right]. 
\]
The precoder set for this example is given by
\[
{\mathbb P}=\left\{ 
\begin{bmatrix} q_{11} & q_{13}\\ q_{21} & q_{23} \end{bmatrix},
\begin{bmatrix} q_{11} & q_{14}\\ q_{21} & q_{24} \end{bmatrix},
\begin{bmatrix} q_{12} & q_{13}\\ q_{22} & q_{23} \end{bmatrix},
\begin{bmatrix} q_{12} & q_{14}\\ q_{22} & q_{24} \end{bmatrix}
\right\}.
\]

{\em System model:}
The PIM transmitter is shown in Fig. \ref{pimsys}. It takes 
$\lfloor\log_2|{\mathbb P}|\rfloor + p\lfloor\log_2|\sa|\rfloor$
bits and encodes them as follows. The $p\lfloor\log_2|\sa|\rfloor$ 
bits are used to obtain $p$ modulation symbols. Let $\vx \in \sa^p$
denote the vector of these modulation symbols. The vector $\vx$ is
precoded by a precoder matrix $\pp_j$ chosen from $\mathbb P$ whose 
index is given by the $\lfloor\log_2|{\mathbb P}|\rfloor$ bits. The 
transmitter then sends one precoded symbol in every channel use. The 
detection is performed after $p$ channel uses. The $p\times 1$ received 
signal vector $\vy$ in this system model can be written as 
\begin{equation}
\vy=\md\pp_j\vx+\vn,
\label{pim_sys}
\end{equation}
where $\pp_j$ is the $p\times p$ PRPP matrix chosen from $\mathbb P$, 
and $\md$ is the channel matrix as described in Sec. \ref{sec3}. The 
ML detection rule is given by
\begin{equation} 
\label{mlpim} 
\{\hat{\vx},\hat{j}\}=\argmin_{\vx\in \sa^p,\, j=1,\cdots,|{\mathbb P}|} \ \|\vy-\md\pp_j\vx\|^2.  
\end{equation}

{\em PIM with activation pattern matrix:} The PIM scheme can also 
be described as follows.
Consider an activation pattern matrix $\mb$ generated using 
$\lfloor\log_2|{\mathbb P}|\rfloor=p\lfloor\log_2n_p\rfloor$
bits, similar to the activation pattern matrix described in Sec. 
\ref{sec2c}. Note that $\mb\vx\in\sv^p$. The $pn_p\times 1$ vector
$\mb\vx$ is precoded with the matrix $\pq$ to get the $p\times 1$
transmit vector. The received signal vector in $p$ channel uses can 
be expressed as
\begin{equation}
\vy=\md\pq\mb\vx+\vn.
\label{vsm_sys}
\end{equation}
The matrix $\pq\mb$ is a $p \times p$ PRPP matrix. The matrix $\mb$
chooses one column from the $n_p$ columns of every $\pq_i$ to get a PRPP 
matrix $\pp_j$. Thus, the choice of $p$ from $pn_p$ columns
conveys $p\lfloor\log_2n_p\rfloor$ information bits, where $n_p$ can be 
viewed as the precoder equivalent of the spatial antennas in SM, which are 
indexed by the information bits. The ML detection rule for this alternate
system model is given by
\begin{equation} 
\label{ml4} 
\{\hat{\vx},\hat{\mb}\}=\argmin_{\vx\in \sa^p, \ \mb} \ \|\vy-\md\pq\mb\vx\|^2.
\end{equation}
The indices of the non-zero rows in $\hat{\mb}$ and the entries of 
$\hat{\mathbf x}$ are demapped to obtain the information bits.
The precoder matrices convey information bits in addition to providing 
diversity, and we can achieve the advantages of SM-MIMO in a SISO fading 
channel through PIM. Also, by taking the view of (\ref{vsm_sys}), detection
algorithms meant for SM can be effectively used for PIM detection.

\begin{figure}
\includegraphics[height=2.75in,width=3.75in]{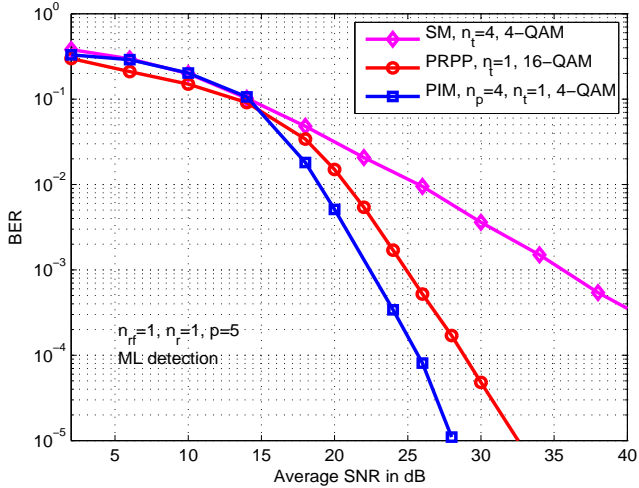}
\caption{Comparison of the BER performance of PIM, SM without PRPP, and
PRPP without SM. $p=5$, 4 bpcu, and ML detection. }
\vspace{-4mm}
\label{pimsim}
\end{figure}

{\em Simulation results}: In Fig. \ref{pimsim}, we present the BER performance 
of the PIM scheme with $n_t=1, n_{rf}=1, n_p=4, n_r=1, p=5$ and 4-QAM, using
ML detection. We compare this performance with the ML detection performance of 
PRPP scheme without SM for $n_t=1, n_{rf}=1, n_r=1, p=5$ and 16-QAM, and 
SM scheme without PRPP for $n_t=4, n_{rf}=1, n_r=1$ and 4-QAM. Note that 
the spectral efficiency in all the three schemes is 4 bpcu. We see that PIM 
outperforms PRPP without SM and SM without PRPP. For e.g., at $10^{-3}$ 
BER, PIM preforms better than PRPP without SM by about 2.5 dB and better 
than SM 
without PRPP by about 12.5 dB. The PIM scheme achieves better performance 
than PRPP without SM because PIM can use a smaller sized (and hence more
power efficient) modulation alphabet compared to PRPP without SM. Also, the
reason behind the better performance of PIM compared to SM without PRPP 
is that PIM provides diversity gain due to precoding. Thus, PIM provides 
the benefits of both diversity advantage of PRPP and SNR advantage of SM. 
Note that the PRPP-SM scheme in Sec. \ref{sec2c} also provides both these 
advantages, but the possibility of using smaller-sized modulation alphabet 
in PRPP-SM arises due to antenna indexing, whereas in PIM it arises due 
to precoder indexing. Therefore, PIM avoids the need to use multiple transmit 
antennas compared to PRPP-SM. This observation leads us to consider
exploiting the antenna indexing in SM for further reduction in modulation
alphabet size in PIM. We refer to such a PIM scheme that exploits both 
precoder indexing as well as antenna indexing as PIM-SM scheme. The
proposed PIM-SM scheme is presented in the following subsection. 

\begin{table*}
\centering
\resizebox{3.6in}{!} {
\begin{tabular}{|c||c|c|c||c|c|}
\hline
 Scheme &\multicolumn{3}{|c||}{Number of bits conveyed through}
		   &\multicolumn{2}{|c|}{Precoding done on}\\
\cline{2-6}\cline{2-6}
&Modulation &Precoder &Antenna &Modulation & Antenna\\
& symbols & index & index & symbols & index bits\\
\hline\hline
PRPP&  	$\lfloor\log_2\sa\rfloor$& - & - & yes & - \\ \hline
SM&  	$\lfloor\log_2\sa\rfloor$& - & $\lfloor\log_2n_t\rfloor$ & no & no \\ \hline
PRPP-SM&$\lfloor\log_2\sa\rfloor$& - & $\lfloor\log_2n_t\rfloor$ & yes & yes \\ \hline
PIM&	$\lfloor\log_2\sa\rfloor$& $\lfloor\log_2n_p\rfloor$ & - & yes & - \\ \hline
PIM-SM&	$\lfloor\log_2\sa\rfloor$& $\lfloor\log_2n_p\rfloor$ & $\lfloor\log_2n_t\rfloor$ & yes & yes \\ \hline
\end{tabular} }
\vspace{1mm}
\caption{Modes of information bits conveyed and precoding done in different 
schemes.}
\label{tab}
\vspace{-6mm}
\end{table*}

\subsection{Precoder index modulation with SM (PIM-SM)}
\label{sec3b}
The proposed PIM-SM scheme uses $n_t$ transmit antennas and one transmit
RF chain, so that $\lfloor\log_2n_t\rfloor$ bits are conveyed as antenna 
index bits. These bits are in addition to the 
$\lfloor\log_2n_p\rfloor+\lfloor\log_2|\sa|\rfloor$ bits conveyed in
PIM. Therefore, the spectral efficiency of the PIM-SM scheme is 
\[
\lfloor\log_2n_p\rfloor+\lfloor\log_2n_t\rfloor+\lfloor\log_2|\sa|\rfloor
\quad \mbox{bpcu}.
\]
For immediate reference and comparison, the spectral efficiencies achieved 
by the different schemes and the bits that are precoded in these schemes 
are tabulated in Table \ref{tab}.

{\em System model:}
The PIM-SM transmitter is illustrated in Fig. \ref{pimsmsys}. The system 
model for the PIM-SM scheme can be written as 
\begin{equation}
\vy=\md\ma\pp\ma\pp_j\vx+\vn,
\label{psim_sys}
\end{equation}
where $\ma$ is the activation pattern matrix of the PRPP-SM scheme 
defined in Sec. \ref{sec2c}.
The ML detection rule for PIM-SM can then be written as 
\begin{equation} 
\label{ml5} 
\{\hat{\vx},\hat{\ma},\hat{j}\}=\argmin_{\vx\in \sa^p, \ma, j=1,\cdots,|{\mathbb P}|} \ \|\vy-\md\ma\pp\ma\pp_j\vx\|^2.  
\end{equation}
The indices of the non-zero rows in $\hat{\ma}$, index $\hat{j}$, and the 
entries of $\hat{\mathbf x}$ are demapped to obtain the information bits.

\begin{figure}
\includegraphics[height=2.75in,width=3.75in]{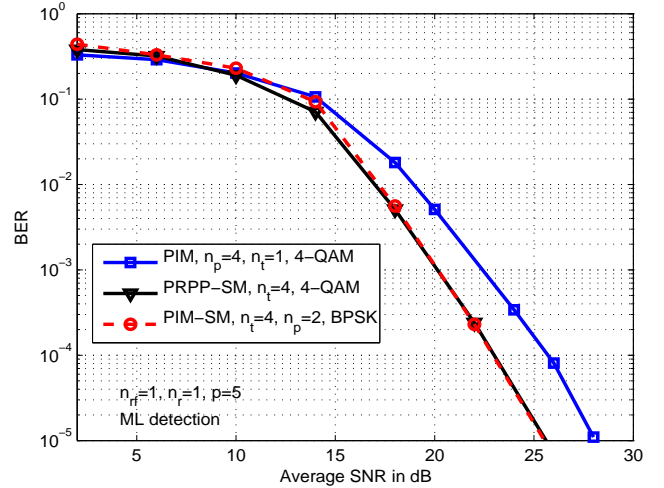}
\caption{Comparison of the BER performance of PIM, PRPP-SM, and PIM-SM 
schemes. $p=5$, 4bpcu, and ML detection. }
\vspace{-2mm}
\label{psim}
\end{figure}

{\em Simulation results}:
In Fig. \ref{psim}, we present the BER performance of PIM-SM scheme with
$n_t=4, n_{rf}=1, n_r=1, n_p=2$, BPSK and $p=5$, using ML detection. 
We also plot the ML detection performance of the PIM scheme (without SM) 
for $n_t=1, n_{rf}=1, n_r=1, n_p=4$, 4-QAM, and $p=5$. We have also
plotted the ML detection performance of PRPP-SM scheme with 
$n_t=4, n_{rf}=1, n_r=1$, 4-QAM, and $p=5$. Note that the spectral 
efficiency is 4 bpcu in all the three schemes. It can be seen that the 
PIM-SM scheme performs better than PIM scheme. This is because of the
smaller-sized modulation alphabet in PIM-SM (BPSK) compared that in PIM 
(4-QAM). Also, PIM-SM performance is similar to that of PRPP-SM. 

\section{Conclusions}
\label{sec4}
We introduced precoder index modulation (PIM) in this paper. We proposed 
a PIM scheme which conveys additional information bits through the choice 
of a precoding matrix from a set of pre-determined PRPP matrices. Combining 
the PIM and PRPP-SM schemes, we proposed a PIM-SM scheme which conveys bits 
through both antenna index as well as precoder index. We showed the 
performance of PIM and PIM-SM schemes for small precoder sizes using ML 
detection. Design of low complexity detection algorithms for PIM and PIM-SM 
for large precoder sizes is open for future extension.

\bibliographystyle{ieeetr}

\begin{thebibliography}{99}
\bibitem{proakis}
J. G. Proakis, {\em Digital Communications}, 4th Ed., Mc-Graw Hill, 2001.

\bibitem{sm}
M. Di Renzo, H. Haas, A. Ghrayeb, S. Sugiura, and L. Hanzo, ``Spatial
modulation for generalized MIMO: challenges, opportunities and 
implementation,'' {\em Proceedings of the IEEE}, vol. 102, no. 1, 
pp. 53-55, Jan. 2014. 

\bibitem{book}
A. Chockalingam and B. Sundar Rajan, {\em Large MIMO Systems}, Cambridge
Univ. Press, Feb. 2014.

\bibitem{sub_index1}
R. Abu-alhiga and H. Haas, ``Subcarrier-index modulation OFDM,''
{\em Proc. IEEE PIMRC'2009}, pp. 177-181, Sep. 2009.

\bibitem{sub_index2}
E. Basar, U. Aygolu, E. Panayirci, and H. V. Poor, ``Orthogonal frequency
division multiplexing with index modulation,'' {\em Proc. IEEE GLOBECOM'2012},
pp. 4741-4746, Dec. 2012.

\bibitem{marco}
N. Serafimovski1, S. Sinanovic, M. Di Renzo, and H. Haas, ``Multiple access
spatial modulation,'' {\em EURASIP J. Wireless Commun. and Networking 2012},
2012:299.

\bibitem{tln}
T. Lakshmi Narasimhan, P. Raviteja, and A. Chockalingam, ``Large-scale
multiuser SM-MIMO versus massive MIMO,'' {\em IEEE ITA'2014}, San Diego, 
Feb. 2014. Also available online: arXiv:1311.1291 [cs.IT] 6 Nov 2013.

\bibitem{tse}
D. Tse and P. Viswanath, {\em Fundamentals of Wireless Communication},
Cambridge University Press, 2005.

\bibitem{vinod}
V. Sharma, K. Premkumar and R. N. Swamy, ``Exponential diversity achieving 
spatio-temporal power allocation scheme for fading channels,'' {\em IEEE 
Trans. Info. Theory}, vol. 54, no. 1, pp. 188-208, Jan. 2008.

\bibitem{ramesh}
R. Annavajjala and P. V. Orlik, ``Achieving near exponential diversity on 
uncoded low-dimensional MIMO, multi-user and multi-carrier systems without 
transmitter CSI,'' {\em Proc. ITA 2011}, Jan. 2011.

\bibitem{lmimo1}
K. V. Vardhan, S. K. Mohammed, A. Chockalingam, and B. S. Rajan, 
``A low-complexity detector for large MIMO systems and multicarrier CDMA 
systems,'' {\em IEEE J. Sel. Areas Commun.,} vol. 26, no. 3, pp. 473-485,
Apr. 2008.

\bibitem{prpp_sm1}
T. Lakshmi Narasimhan, Y. Naresh, T. Datta, and A. Chockalingam,
``Pseudo-random phase precoded spatial modulation,''
available online arXiv:1401.6543v1 [cs.IT] 25 Jan 2014.

\bibitem{prpp_sm2}
T. Lakshmi Narasimhan, Y. Naresh, T. Datta, and A. Chockalingam,
``Pseudo-random phase precoded spatial modulation and precoder index
modulation,'' accepted in {\em IEEE GLOBECOM'2014}, Dec. 2014.

\end{thebibliography}

\end{document}